\documentclass[12pt]{article}
\usepackage[dvips]{graphicx}
\usepackage{pdproc}
\makeatletter 
\def\@cite#1{[#1]} 
\makeatother    
\begin{document}

\newcommand {\nn}{\nonumber \\}
\newcommand {\Tr}{{\rm Tr\,}}
\newcommand {\tr}{{\rm tr\,}}
\newcommand {\e}{{\rm e}}
\newcommand {\etal}{{\it et al.}}
\newcommand {\m}{\mu}
\newcommand {\n}{\nu}
\newcommand {\pl}{\partial}
\newcommand {\p} {\phi}
\newcommand {\vp}{\varphi}
\newcommand {\vpc}{\varphi_c}
\newcommand {\al}{\alpha}
\newcommand {\be}{\beta}
\newcommand {\ga}{\gamma}
\newcommand {\Ga}{\Gamma}
\newcommand {\x}{\xi}
\newcommand {\ka}{\kappa}
\newcommand {\la}{\lambda}
\newcommand {\La}{\Lambda}
\newcommand {\si}{\sigma}
\newcommand {\sh}{\theta}
\newcommand {\Th}{\Theta}
\newcommand {\om}{\omega}
\newcommand {\Om}{\Omega}
\newcommand {\ep}{\epsilon}
\newcommand {\vep}{\varepsilon}
\newcommand {\na}{\nabla}
\newcommand {\del}  {\delta}
\newcommand {\Del}  {\Delta}
\newcommand {\mn}{{\mu\nu}}
\newcommand {\ls}   {{\lambda\sigma}}
\newcommand {\ab}   {{\alpha\beta}}
\newcommand {\half}{ {\frac{1}{2}} }
\newcommand {\third}{ {\frac{1}{3}} }
\newcommand {\fourth} {\frac{1}{4} }
\newcommand {\sixth} {\frac{1}{6} }
\newcommand {\sqg} {\sqrt{g}}
\newcommand {\sqtwo} {\sqrt{2}}
\newcommand {\fg}  {\sqrt[4]{g}}
\newcommand {\invfg}  {\frac{1}{\sqrt[4]{g}}}
\newcommand {\sqZ} {\sqrt{Z}}
\newcommand {\gbar}{\bar{g}}
\newcommand {\sqk} {\sqrt{\kappa}}
\newcommand {\sqt} {\sqrt{t}}
\newcommand {\reg} {\frac{1}{\epsilon}}
\newcommand {\fpisq} {(4\pi)^2}
\newcommand {\Lcal}{{\cal L}}
\newcommand {\Ocal}{{\cal O}}
\newcommand {\Dcal}{{\cal D}}
\newcommand {\Ncal}{{\cal N}}
\newcommand {\Mcal}{{\cal M}}
\newcommand {\scal}{{\cal s}}
\newcommand {\Dvec}{{\hat D}}   
\newcommand {\dvec}{{\vec d}}
\newcommand {\Evec}{{\vec E}}
\newcommand {\Hvec}{{\vec H}}
\newcommand {\Vvec}{{\vec V}}
\newcommand {\Btil}{{\tilde B}}
\newcommand {\ctil}{{\tilde c}}
\newcommand {\Ftil}{{\tilde F}}
\newcommand {\Stil}{{\tilde S}}
\newcommand {\Ztil}{{\tilde Z}}
\newcommand {\Ktil}  {{\tilde K}}
\newcommand {\Ltil}  {{\tilde L}}
\newcommand {\Qtil}  {{\tilde Q}}
\newcommand {\altil}{{\tilde \alpha}}
\newcommand {\betil}{{\tilde \beta}}
\newcommand {\latil}{{\tilde \lambda}}
\newcommand {\ptil}{{\tilde \phi}}
\newcommand {\Ptil}{{\tilde \Phi}}
\newcommand {\natil} {{\tilde \nabla}}
\newcommand {\ttil} {{\tilde t}}
\newcommand {\Rhat}{{\hat R}}
\newcommand {\Shat}{{\hat S}}
\newcommand {\shat}{{\hat s}}
\newcommand {\Dhat}{{\hat D}}   
\newcommand {\Vhat}{{\hat V}}   
\newcommand {\xhat}{{\hat x}}
\newcommand {\Zhat}{{\hat Z}}
\newcommand {\Gahat}{{\hat \Gamma}}
\newcommand {\nah} {{\hat \nabla}}
\newcommand {\gh}  {{\hat g}}
\newcommand {\labar}{{\bar \lambda}}
\newcommand {\cbar}{{\bar c}}
\newcommand {\bbar}{{\bar b}}
\newcommand {\Bbar}{{\bar B}}
\newcommand {\Dbar}{{\bar D}}
\newcommand {\Qbar}{{\bar Q}}
\newcommand {\Kbar}  {{\bar K}}
\newcommand {\Lbar}  {{\bar L}}
\newcommand {\albar}{{\bar \alpha}}
\newcommand {\psibar}{{\bar \psi}}
\newcommand {\Psibar}{{\bar \Psi}}
\newcommand {\sibar}{{\bar \si}}
\newcommand {\thbar}{{\bar \theta}}
\newcommand {\chibar}{{\bar \chi}}
\newcommand {\xibar}{{\bar \xi}}
\newcommand {\bbartil}{{\tilde {\bar b}}}
\newcommand {\aldot} {{\dot \al}}
\newcommand {\bedot} {{\dot \be}}
\newcommand {\deldot} {{\dot \delta}}
\newcommand {\gadot} {{\dot \ga}}
\newcommand  {\vz}{{v_0}}
\newcommand  {\ez}{{e_0}}
\newcommand {\intfx} {{\int d^4x}}
\newcommand {\inttx} {{\int d^2x}}
\newcommand {\change} {\leftrightarrow}
\newcommand {\ra} {\rightarrow}
\newcommand {\larrow} {\leftarrow}
\newcommand {\ul}   {\underline}
\newcommand {\pr}   {{\quad .}}
\newcommand {\com}  {{\quad ,}}
\newcommand {\q}    {\quad}
\newcommand {\qq}   {\quad\quad}
\newcommand {\qqq}   {\quad\quad\quad}
\newcommand {\qqqq}   {\quad\quad\quad\quad}
\newcommand {\qqqqq}   {\quad\quad\quad\quad\quad}
\newcommand {\qqqqqq}   {\quad\quad\quad\quad\quad\quad}
\newcommand {\qqqqqqq}   {\quad\quad\quad\quad\quad\quad\quad}
\newcommand {\lb}    {\linebreak}
\newcommand {\nl}    {\newline}

\newcommand {\vs}[1]  { \vspace*{#1 mm} }

\newcommand {\MPL}  {Mod.Phys.Lett.}
\newcommand {\IJMP}  {Int.Jour.Mod.Phys.}
\newcommand {\NP}   {Nucl.Phys.}
\newcommand {\PL}   {Phys.Lett.}
\newcommand {\PR}   {Phys.Rev.}
\newcommand {\PRL}   {Phys.Rev.Lett.}
\newcommand {\CMP}  {Commun.Math.Phys.}
\newcommand {\JMP}  {Jour.Math.Phys.}
\newcommand {\AP}   {Ann.of Phys.}
\newcommand {\PTP}  {Prog.Theor.Phys.}
\newcommand {\NC}   {Nuovo Cim.}
\newcommand {\CQG}  {Class.Quantum.Grav.}

\def\graph#1{
\begin{array}{c}\mbox{
\includegraphics*[height=1cm]{#1.eps}
}\end{array}
             }

\font\smallr=cmr5
\def\ocirc#1{#1^{^{{\hbox{\smallr\llap{o}}}}}}
\def\ogamma{\ocirc{\gamma}{}}
\def\oM{{\buildrel {\hbox{\smallr{o}}} \over M}}
\def\osigma{\ocirc{\sigma}{}}

\def\overleftrightarrow#1{\vbox{\ialign{##\crcr
 $\leftrightarrow$\crcr\noalign{\kern-1pt\nointerlineskip}
 $\hfil\displaystyle{#1}\hfil$\crcr}}}
\def\overnab{{\overleftrightarrow\nabslash}}

\def\va{{a}}
\def\vb{{b}}
\def\vc{{c}}
\def\tilpsi{{\tilde\psi}}
\def\tbpsi{{\tilde{\bar\psi}}}



\title{
 Graphical Representation of Supersymmetry
 and Computer Calculation
}

\author{ SHOICHI ICHINOSE}

\address{ 
School of Food and Nutritional Sciences, University of Shizuoka\\
Yada 52-1, Shizuoka 422-8526, Japan
\\ {\rm E-mail: ichinose@u-shizuoka-ken.ac.jp}}

\abstract{
A graphical representation of supersymmetry
is presented. It clearly expresses the chiral flow 
appearing in SUSY quantities, by representing
spinors by {\it directed lines} (arrows). The chiral suffixes
are expressed by the directions (up, down, left, right)
of the arrows. 
The SL(2,C) invariants are represented by {\it wedges}. 
We are free from the messy symbols of spinor suffixes.
The method is applied to the 5D supersymmetry. 
Many applications are expected. 
The result is suitable for coding a computer program
and is highly expected to be applicable to 
various SUSY theories (including Supergravity) 
in various dimensions. 
}

\normalsize\baselineskip=15pt

\vspace{5mm}

{\it 1. Introduction}\ \ 
The beauty of Supersymmetry(SUSY) 
comes from the harmony between bosons and fermions. 
At the cost of the high symmetry, the SUSY fields
generally carry many suffixes:\ 
chiral-suffixes ($\al$), anti-chiral suffixes ($\aldot$)
in addition to usual ones:\ 
gauge suffixes ($i,j,..$), Lorentz suffixes ($m,n,\cdots$).
The usual notation is, for example, $\psi^{i~\aldot}_{m\al}$.
Many suffixes are ``crowded'' within one character $\psi$.
Whether the meaning of a quantity is clearly read, 
sometimes crucially depends on
the way of description. In the case of quantities
with many indices,  
we are sometimes lost in the ``jungle'' of suffixes.
In this circumstance we propose a new representation
to express SUSY quantities[1]
.
It has the following properties:\ 
1. 
All suffix-information is expressed;\ 
2. 
Suffixes are suppressed as much as possible. 
Instead we use the ``geometrical" notation:\ lines, arrows, $\cdots$.
Particularly, 
contracted suffixes (we call them ``dummy'' suffixes) are
expressed by vertices (for fermion suffixes) or lines;\ 
3. 
The chiral flow is manifest;\ 
4. 
The {\it graphical indices} (defined in Sect.4) specify a spinorial quantity.

Another quantity with many suffixes
is the Riemann tensor appearing in the general relativity.
It was already graphically represented[2] 
and some applications have appeared[3,4]
. 
We follow the notation of the textbook by
Wess and Bagger[5]
.

{\it 2. Definitions}\ \ 
Let us represent the Weyl fermion $\psi_\al,\psibar_\aldot$
(2 complex components, $\al,\aldot=1,2,$ ) 
and their ``suffixes-raised" partners as in Fig.1.
Raising and lowering the spinor suffixes is done by
the antisymmetric tensors $\ep^\ab, \ep_{\ab}$.
\begin{figure}[htb]
\begin{center}
\includegraphics*[width=15cm]{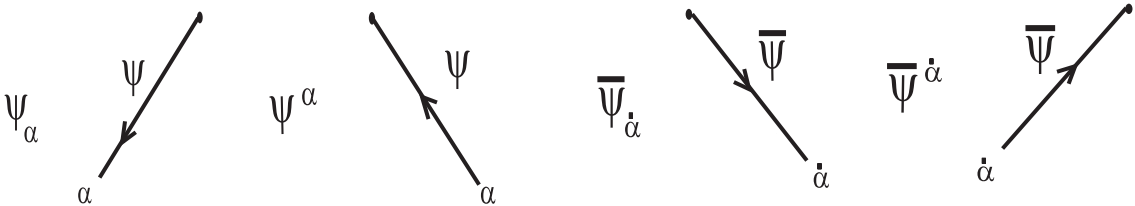}
\caption{%
Weyl fermions.
}
\label{F1}
\end{center}
\end{figure}

We introduce a rule:\ every spinor graph is {\it anticommuting} 
in the horizontal direction.

The elements of the SL(2,C) $\si$-matrix are expressed as in Fig.2. 
\begin{figure}[htb]
\begin{center}
\includegraphics*[width=12cm]{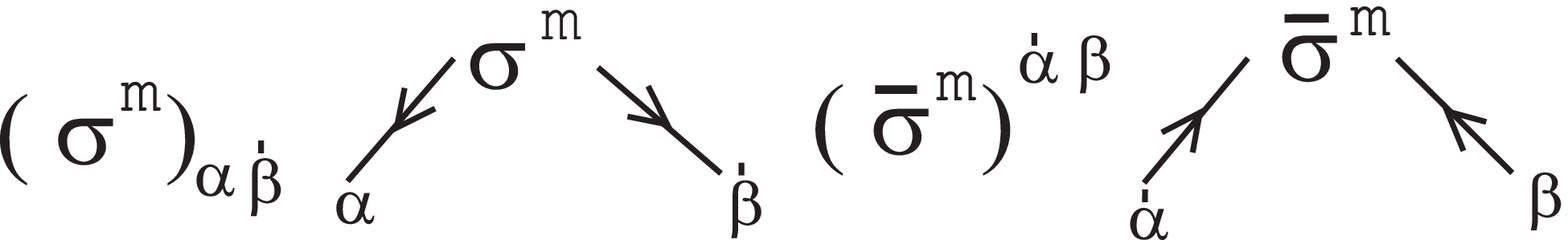}
\caption{%
Elements of SL(2,C) $\si$-matrices. 
$(\si^m)_{\al\bedot}$ and $(\sibar^m)^{\aldot\be}$ are
the standard form.
}
\label{F3}
\end{center}
\end{figure}

Two Lorentz vectors,
$\chi^\al (\si^m)_{\al\bedot}\psibar^\bedot$ and
$\psibar_\aldot(\sibar^m)^{\aldot\be}\chi_\be$
are expressed as in Fig.3. 

\begin{figure}[htb]
\begin{center}
\includegraphics*[width=10cm]{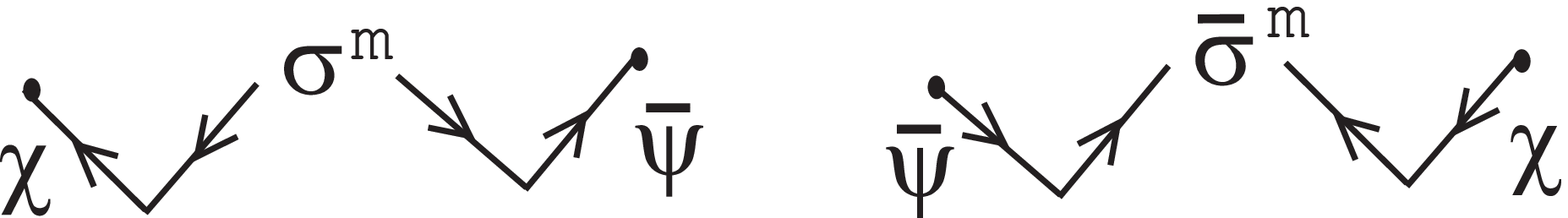}
\caption{%
Two Lorentz vectors 
$\chi^\al (\si^m)_{\al\bedot}\psibar^\bedot$ and
$\psibar_\aldot(\sibar^m)^{\aldot\be}\chi_\be$. 
The double wedge structure appears. 
}
\label{F5}
\end{center}
\end{figure}

We can express all formulae graphically.
We here list only basic ones.
In Fig.4, the symmetric combination of $\sibar^m\si^n$
are shown as the basic spinor algebra.

\begin{figure}[htb]
\begin{center}
\includegraphics*[width=13cm]{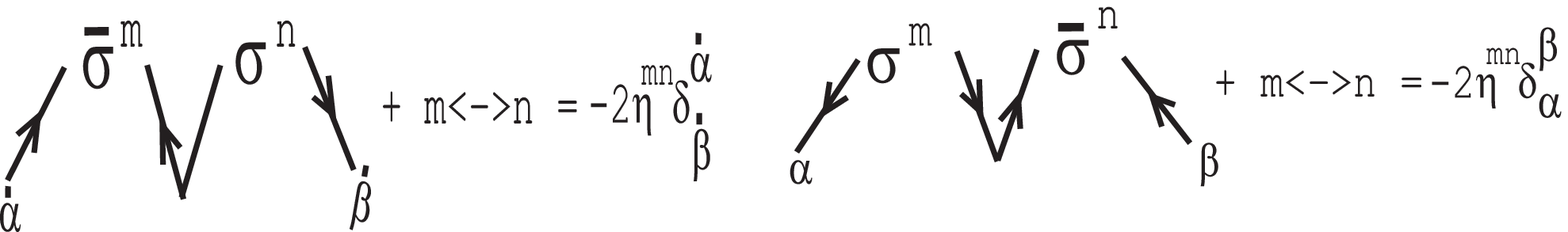}
\caption{%
A graph for  
$(\sibar^m)^{\aldot\be}(\si^n)_{\be\bedot}+m\change n
=-2\eta^{mn}\del^\aldot_\bedot$, and 
$(\si^m)_{\al\aldot}(\sibar^n)^{\aldot\be}+m\change n
=-2\eta^{mn}\del^\be_\al$. 
}
\label{F10}
\end{center}
\end{figure}

The {\it completeness} relations are expressed as in Fig.5.

\begin{figure}[htb]
\begin{center}
\includegraphics*[width=13cm]{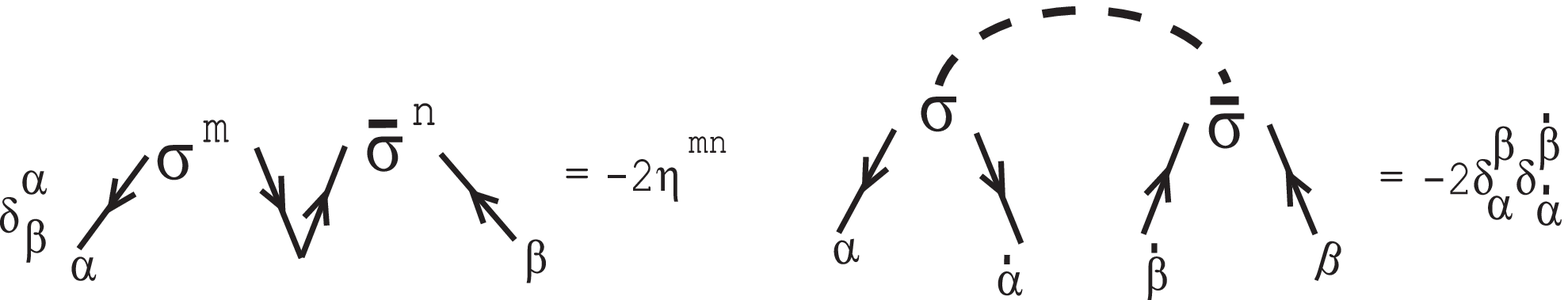}
\caption{%
Completeness relations:
$\del^\al_\be(\si^m)_{\al\aldot}(\sibar^n)^{\aldot\be}
=-2\eta^{mn}$,and  
$(\si^m)_{\al\aldot}(\sibar_m)^{\bedot\be}
=-2\del^\be_\al\del^\bedot_\aldot$.
}
\label{F13}
\end{center}
\end{figure}

We display the Fierz identity .
\begin{eqnarray}
\graph{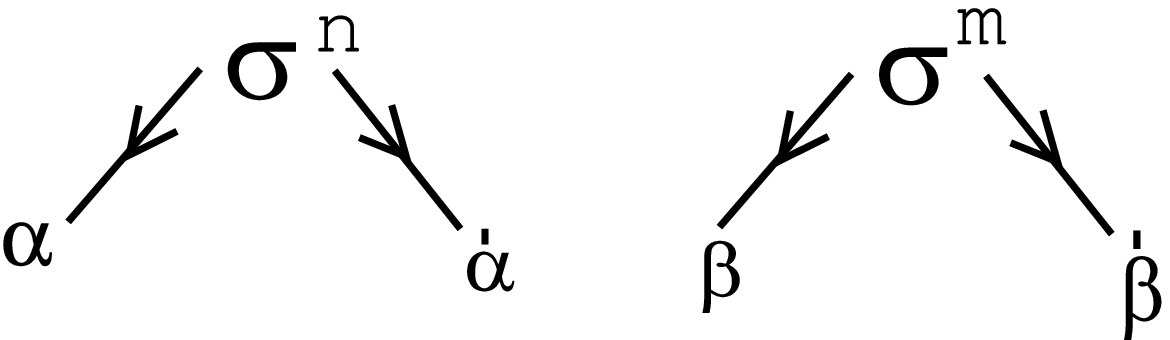}=   
\fourth\left\{
-\graph{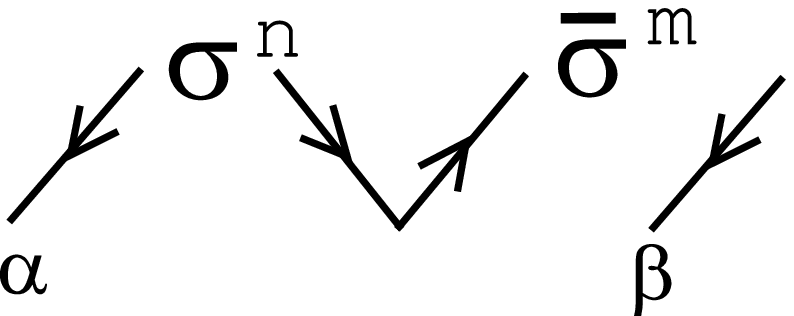}\ep_{\aldot\bedot}+\graph{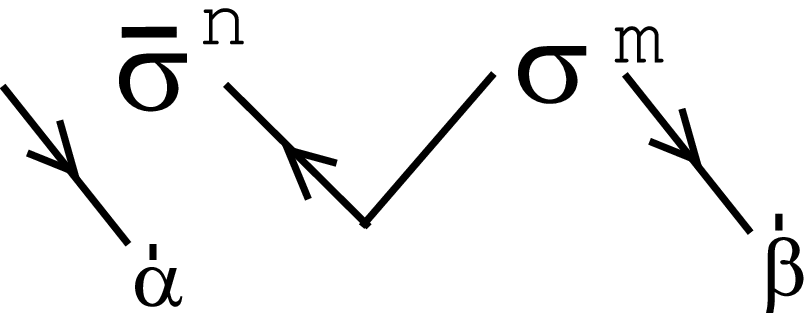}\ep_\ab
-m\change n\right\}\nn
-\half\eta^{nm}\ep_\ab \ep_{\aldot\bedot}
-\frac{1}{8}\left\{\graph{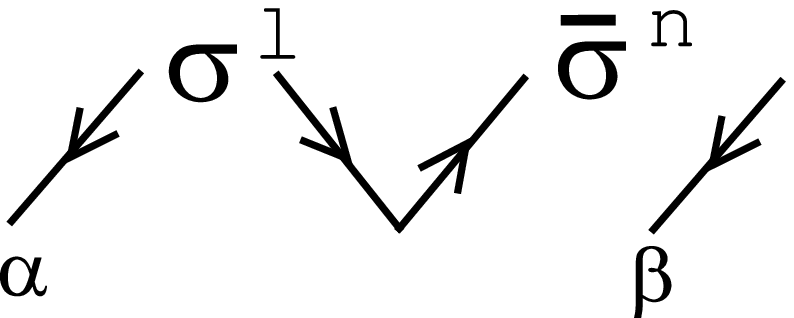}-l\change n\right\}
\left\{\graph{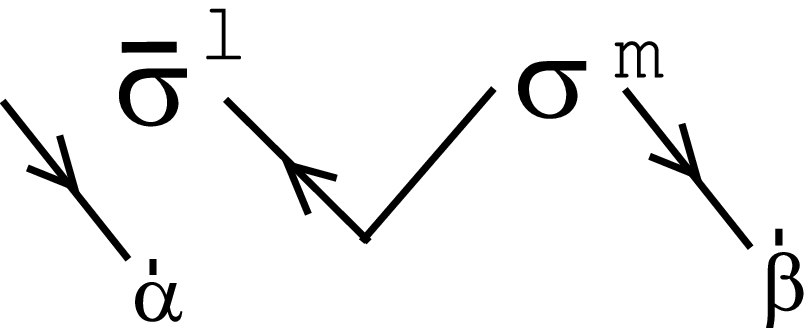}-l\change m\right\}.
\label{def3}
\end{eqnarray}

{\it 3. SUSY Theories and SUSY Transformation}\ \ 
We express the Wess-Zumino model $\Lcal$(
a complex scalar field $A$, a Weyl spinor $\psi_\al$, 
an auxiliary field (complex scalar) $F$) and 
the super elctromagnetic theory in the WZ gauge $\Lcal_{EM}$
($v_{mn}=\pl_mv_n-\pl_nv_m$, 
$v_m$: vector field, $\la$: Weyl fermion, $D$: 
scalar auxiliary field), 
as follows. 
\begin{eqnarray}
\Lcal
=\graph{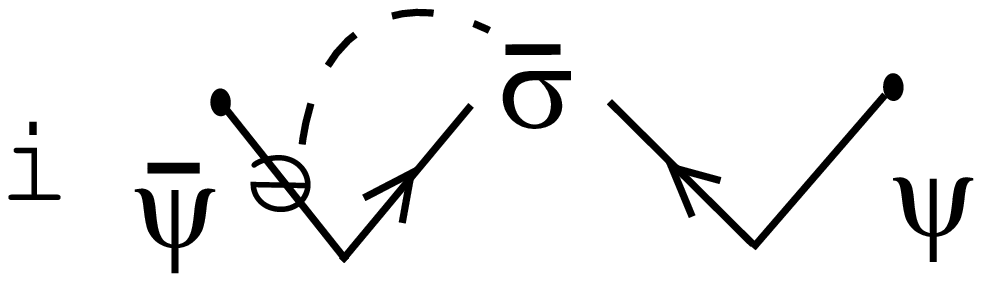}+A^*\pl^2 A+F^*F,
\Lcal_{EM}=\half D^2-\fourth v^{mn}v_{mn}
\graph{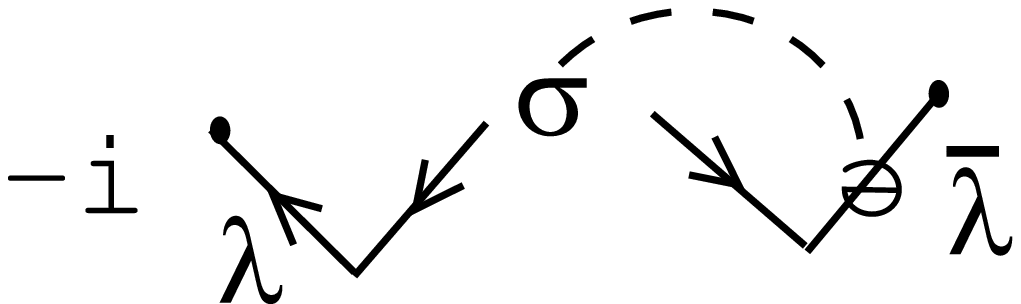}.
\label{chi2}
\end{eqnarray}
The SUSY transformations 
are expressed as
\begin{eqnarray}
\del_\xi A=\sqtwo\graph{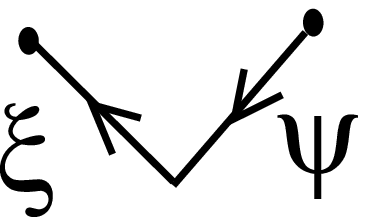}\com\q
\del_\xi F=i\sqtwo\graph{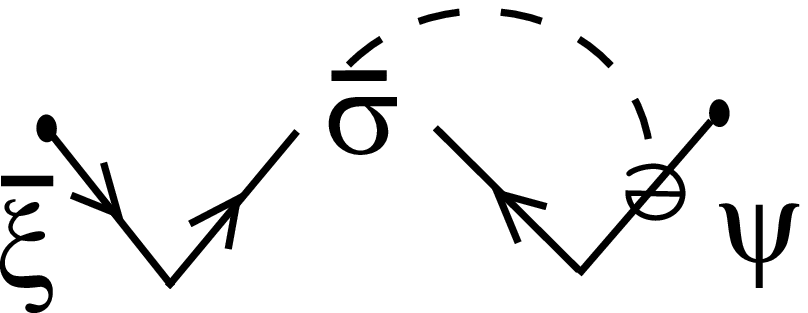}\com\nn
\del_\xi \psi_\al
=i\sqtwo\graph{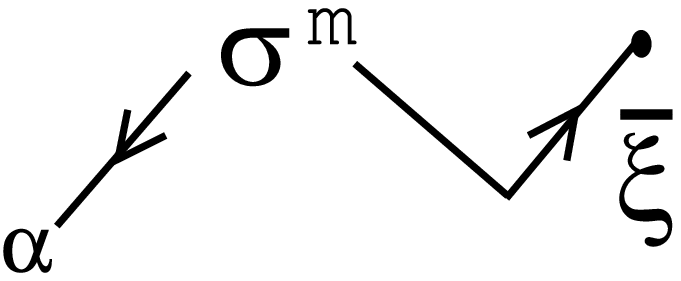}\pl_mA+\sqtwo\graph{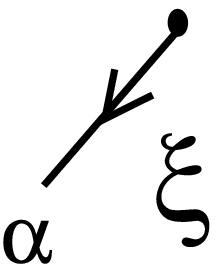}F
\pr\nn
\del_\xi D=
\graph{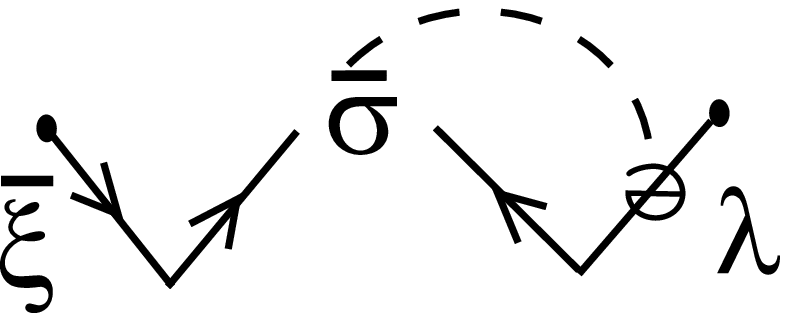} \graph{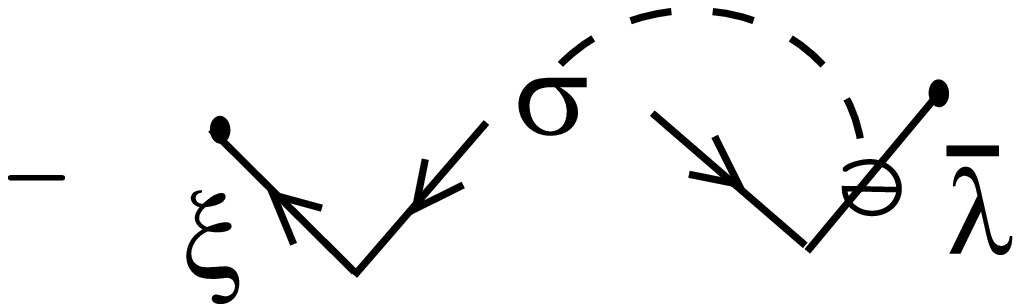}\com\nn
\del_\xi
\graph{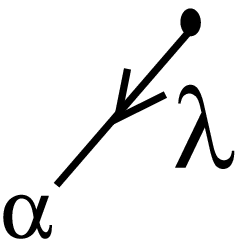}=
\graph{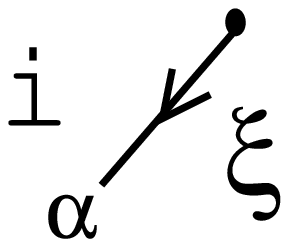}D+\fourth\left\{
\graph{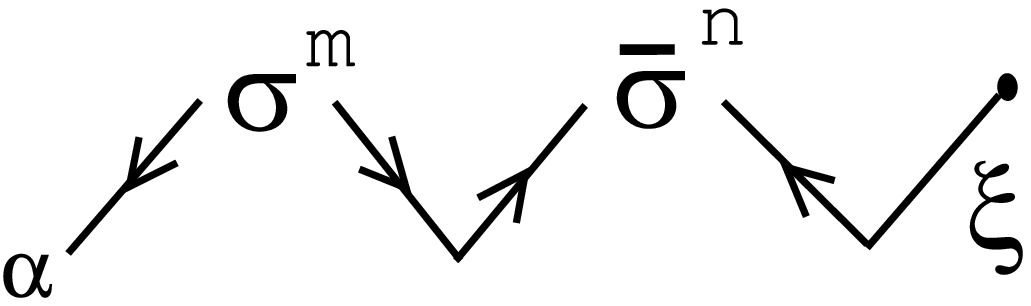}-m\change n
                     \right\}v_{mn}\com\nn
\del_\xi v_{mn}=\graph{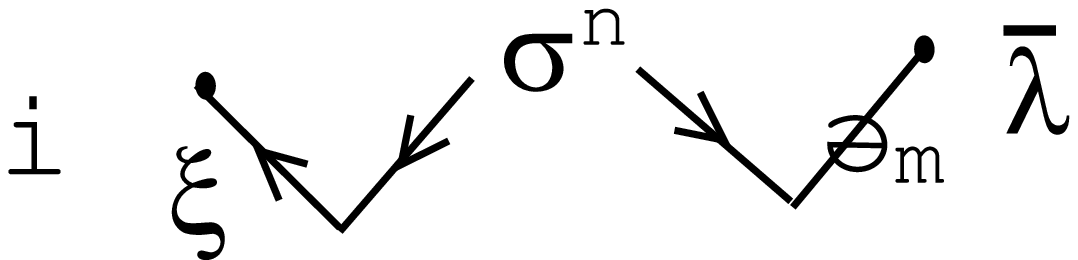} \graph{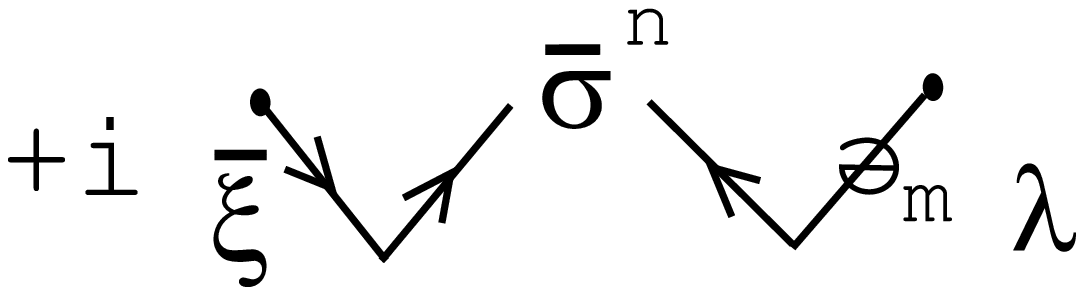}-m\change n\com\nn
\del_\xi v_m=\graph{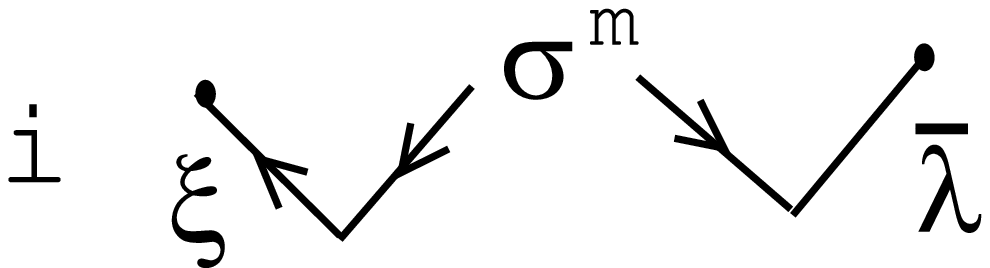} \graph{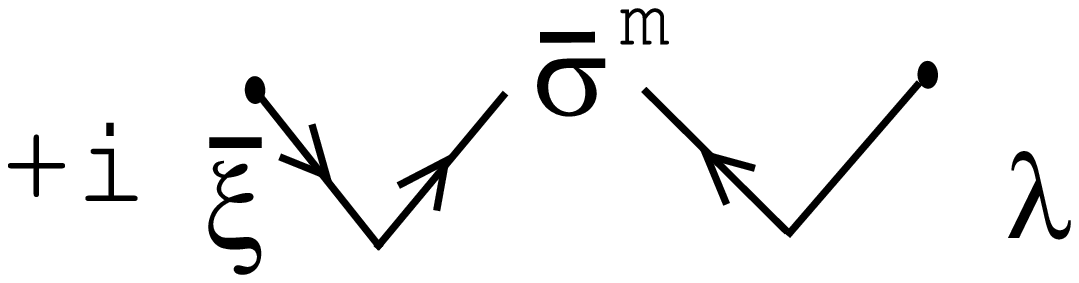}
\pr
\label{chi1a}
\end{eqnarray}
Using the above graphical SUSY transformations, 
we can graphically show the SUSY symmetry of 
$\Lcal$ and $\Lcal_{EM}$ of (\ref{chi2}). 
This shows the correctness of the present representation.

We can read the graphical rule of the complex conjugate
operation by comparing (\ref{chi1a}) and its conjugate.\nl

{\bf Graphical Rule:\ Complex Conjugation Operation}
\begin{eqnarray}
\graph{F1chi1a}\ra\graph{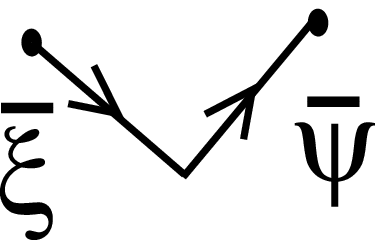}\ ,\ 
\graph{F4chi1a}\ra -\graph{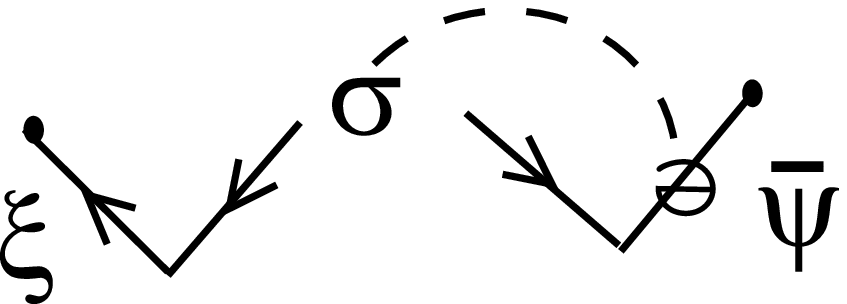}
\ .
\label{chi1c}
\end{eqnarray}

{\it 4. Graph Indices}\ \ 
We introduce some {\it indices} of a graph. They classify graphs.
\nl
(i) {\it Left Chiral Number}(LCN) and {\it Right Chiral Number}(RCN);\ 
(ii) {\it Left Up-Down Number}(LUDN) and {\it Right Up-Down Number}(RUDN) . 
For SL(2,C) invariants, these indices vanish;\ 
(iii) {\it Left Wedge Number}(LWN) and {\it Right Wedge Number}(RWN). 
We assign $LWN=1$ for $\graph{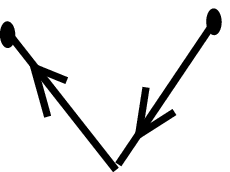}$, and $RWN=1$ for $\graph{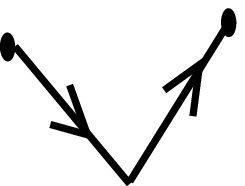}$;\ 
(iv) {\it Dotted Line Number}(DLN).

In addition to the graph-related indices, we introduce\ 
a) Physical Dimension (DIM);\ 
b) Number of the differentials (DIF);\ 
c) Number of $\si$ or $\sibar$ (SIG). 

We list the above indices for basic spinor quantities in Table 1.

\vspace{3mm}

\begin{tabular}{|c|c|c|c|c|}
\hline
             & $\graph{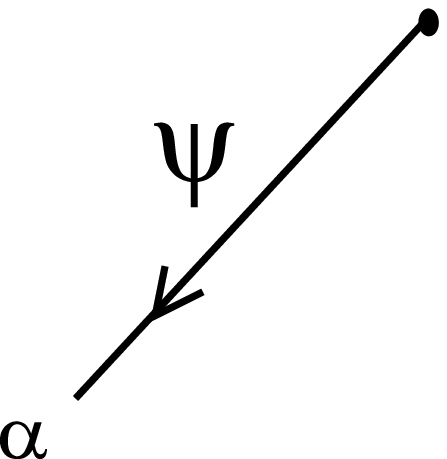}$  & $\graph{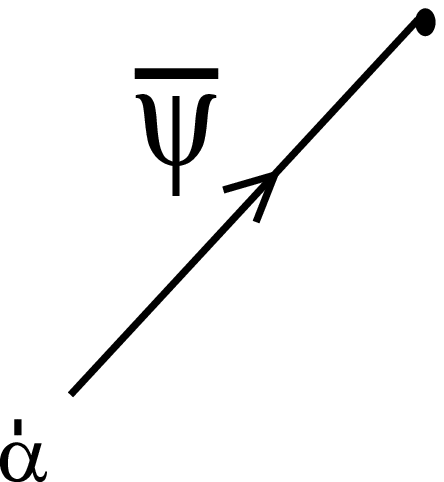}$ & $\graph{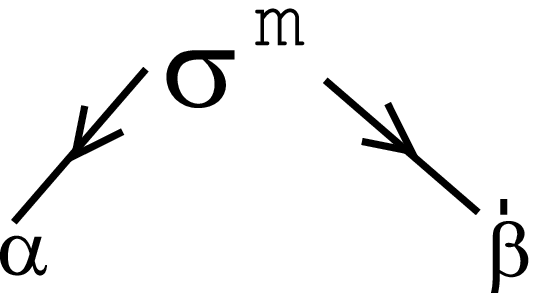}$  & 
                                                   $\graph{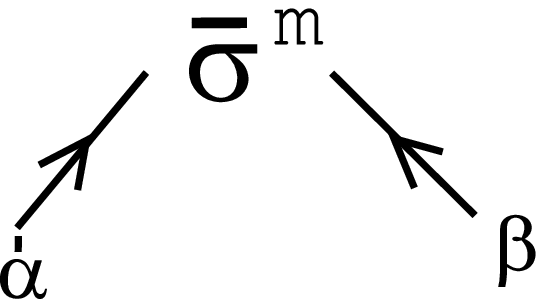}$            \\
\hline
(LCN, RCN)  & $(\half,0)$ & $(0,\half)$ & $(\half,\half)$& $(\half,\half)$\\
(LUDN, RUDN)  & $(-\half,0)$ & $(0,+\half)$ & $(-\half,-\half)$& 
                                                     $(\half,\half)$\\
(LWN, RWN)  &   0         &     0     &    0    &  0      \\
DLN         &  0         &     0     &    0    &  0      \\
DIM         &   $\frac{3}{2}$  &$\frac{3}{2}$  & 0  &  0  \\
DIF     &   0         &     0     &    0    &  0      \\
SIG     &   0     &   0     &   1  &   1  \\
\hline
\multicolumn{5}{c}{\q}                                                 \\
\multicolumn{5}{c}{Table 1\ \ Indices for basic spinor
quantities:\ $\psi_\al, \psibar^\aldot, (\si^m)_{\al\bedot}$
and $(\sibar^m)^{\aldot\be}$.  
}\\
\end{tabular}

{\it 5. Discussion and conclusion}\ \ 
%
As an application of the present approach, supergravity is interesting.
Relegating the full treatment to a separate work, we indicate a graphical
advantage here. There appears such a quantity in the supergravity.:\ 
$\psi_{\del\deldot\ga\gadot\al}=(\si^d)_{\del\deldot}(\si^c)_{\ga\gadot}
e_d^{~n}e_c^{~m}(\psi_{nm})_\al,\ (\psi_{nm})^\al=\pl_n\psi_m^\al+\cdots  
-n\change m.$ Graphically it is expressed as
\begin{eqnarray}
\begin{array}{c}\mbox{
\includegraphics*[height=2cm]{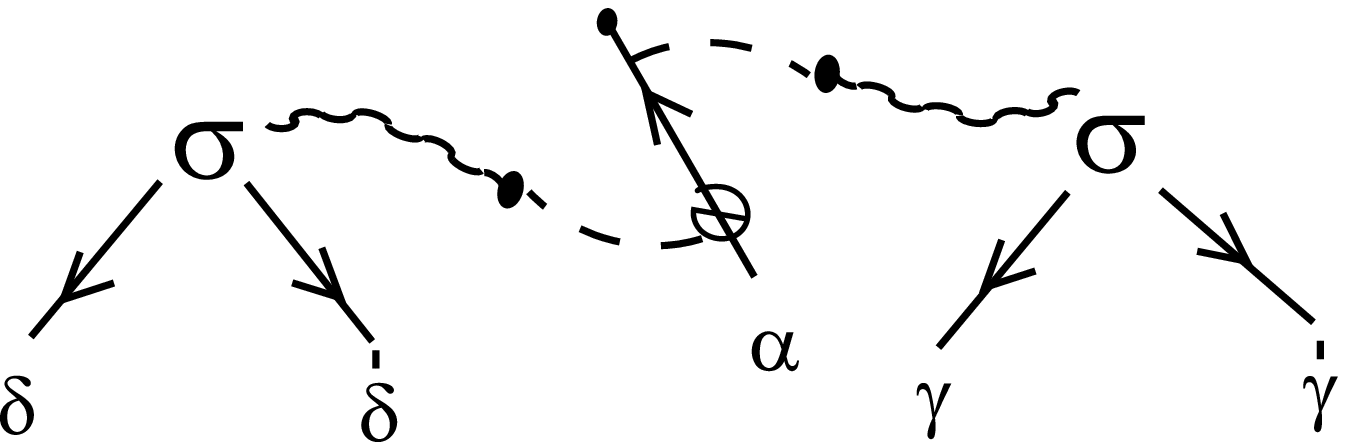}
}\end{array}
\ +\cdots\ ,\ 
e^n_{~a}\ :\ 
\begin{array}{c}\mbox{
\includegraphics*[height=1cm]{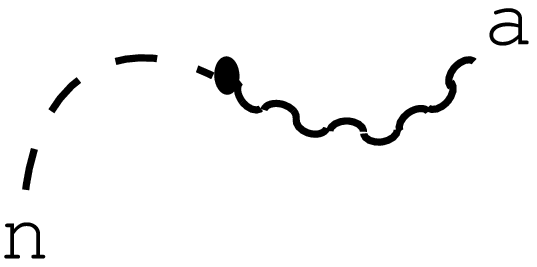}
}\end{array}\ \ 
\psi_m^{~\al}\ :\ 
\begin{array}{c}\mbox{
\includegraphics*[height=1cm]{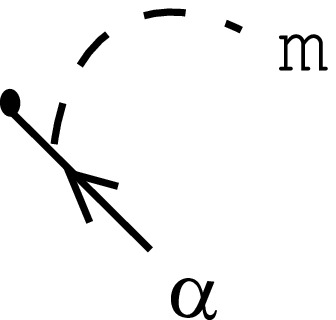}
}\end{array}
\label{conc1}
\end{eqnarray}
where we introduce the graphical representation for the vier-bein $e^n_{~a}$
and the Rarita-Schwinger field $\psi_m^{~\al}$.
The set of indices, which specifies the above graph (\ref{conc1}),
is given as follows:\ 
$(LCN,RCN)=(3/2,1), (LUDN,RUDN)=(-1/2,-1), (LWN,RWN)=(0,0), 
DIM=5/2, DIF=1, SIG=2
$.

%

We have presented a graphical representation
of the supersymmetric theory. It has some
advantages over the conventional description.
The applications are diverse. 
\footnote{
In the conference talk, an application to 5D SUSY is presented.
The progress in the computer calculation is also reported.
See ref.[1]
 for detail.
}
Especially
the higher dimensional suspergravities
are the interesting physical models
to apply the present approach. 
In the ordinary approach, it has a technical problem which
hinders analysis. 
The theory is so "big" that it is rather hard 
in the conventional approach.
The present graphical description is
expected to resolve or reduce the technical
but an important problem. We point out that the
present representation is suitable for coding
as a (algebraic) computer program. 
(See [6,3]
 for the C-language program
and graphical calculation for the product of Riemann tensors.)


{\it 6. References}\ \ \nl
[1]\ S. Ichinose, DAMTP-2003-8, US-03-01, hep-th/0301166,
"Graphical Representation of Supersymmetry"\ 
[2]\ S. Ichinose, \CQG {\bf 12}(1995)1021; hep-th/9309035\ 
[3]\ S. Ichinose and N. Ikeda, \JMP {\bf 38}(1997)6475; hep-th/9702003\ 
[4]\ G.W. Gibbons and S. Ichinose, \CQG {\bf 17}(2000)2129; 
hep-th/9911167\ 
[5]\ J. Wess and J. Bagger, {\it Supersymmetry and Supergravity}. 
Princeton University Press, Princeton, 1992\ 
[6]\ S. Ichinose, \IJMP {\bf C9}(1998)243; hep-th/9609014

\end{document}